\documentclass{aa}
\usepackage{graphicx}
\usepackage{rotating}
\begin{document}

\hyphenation{Fe-bru-ary Gra-na-da mo-le-cu-le mo-le-cu-les}

\title{First detection of triply-deuterated methanol}

\author{
   B. Parise\inst{1}
\and A. Castets\inst{2}
\and E. Herbst\inst{3}
\and E. Caux\inst{1}
\and C. Ceccarelli\inst{4}
\and I. Mukhopadhyay\inst{5}
\and A.G.G.M. Tielens\inst{6}
}
\institute{
CESR CNRS-UPS, BP 4346, 31028 - Toulouse cedex 04, France
\and
Observatoire de Bordeaux, BP 89, 33270 Floirac, France
\and
Department of Physics, The Ohio State University, 174 W. 18th Ave.
Columbus, OH 43210-1106, USA
\and
Laboratoire d'Astrophysique, Observatoire de Grenoble, BP 53, 38041
Grenoble cedex 09, France
\and
Dakota State University, 820 N. Washington Ave., Madison, SD 57042, USA
\and
SRON, P.O. Box 800, NL-9700 AV Groningen, the Netherlands
}
\offprints{Berengere.Parise@cesr.fr}

\date{Received {\today} /Accepted }
\titlerunning{First detection of triply-deuterated methanol}
\authorrunning{Parise et al.}

\abstract{We report the first detection of triply-deuterated methanol, with 
12 observed transitions, towards the low-mass protostar IRAS 16293$-$2422, as well as
multifrequency observations of $^{13}$CH$_3$OH, used to derive the column density of 
the main isotopomer CH$_3$OH.
The derived fractionation ratio [CD$_3$OH]/[CH$_3$OH] averaged on a 10$''$ beam is 1.4\%. 
Together with previous CH$_2$DOH and CHD$_2$OH observations, 
the present CD$_3$OH observations are consistent with a formation of methanol 
on grain surfaces, if the atomic D/H ratio is 0.1 to 0.3 in the accreting 
gas. Such a high atomic ratio can be reached in the frame of gas-phase chemical 
models including all deuterated isotopomers of H$_{3}^{+}$.
 }

\maketitle

\section{Introduction}

Despite the relatively low elemental abundance of deuterium in space 
(a factor of $\sim$~1.5$\times$10$^{-5}$ less abundant than H; \cite{Linsky98} 1998),
extremely large amounts of doubly-deuterated formaldehyde (D$_2$CO/H$_2$CO~$\sim$~10\%)
have been observed in the solar-type protostar IRAS~16293$-$2422 (hereafter 
IRAS~16293, \cite{Ceccarelli98} 1998, \cite{Loinard00} 2000, 
\cite{Ceccarelli01} 2001), initiating the search for other multiply deuterated molecules. 
Subsequently, doubly deuterated formaldehyde, doubly deuterated hydrogen sulfide and 
multiply deuterated ammonia have been observed in other protostars and dark clouds from 
where protostars form (\cite{Roueff00} 2000, \cite{Loinard01} 2001, \cite{Ceccarelli02} 2002, 
\cite{vanderTak02} 2002, \cite{Lis02} 2002, \cite{Vastel03} 2003). 
These studies have been interpreted in terms of two different routes for 
formaldehyde, hydrogen sulfide and ammonia deuteration: active grain chemistry 
followed by at least partial desorption into the gas for 
formaldehyde and hydrogen sulfide on the one hand and gas-phase chemistry for 
ammonia on the other hand. However, ammonia may also be a grain surface product, 
provided a large D/H atomic ratio in the accreting gas.
Recently, doubly-deuterated methanol was detected towards IRAS~16293 
(\cite{Parise02} 2002). 
This observation provided new constraints for chemical models. 
The observations of the deuterated methanols CH$_{2}$DOH and CHD$_{2}$OH were  
both consistent with the formation of methanol from successive 
hydrogenations of CO by reaction with atomic H on grain surfaces, but required  
an atomic D/H ratio of 0.2 to 0.3 in the accreting gas.  At the time of the 
observation of doubly-deuterated methanol, no gas-phase model was able to 
predict such a high atomic D/H ratio. 
Meantime, observations of doubly deuterated formaldehyde in a sample of pre-stellar 
cores showed that the degree of deuteration increases with increasing CO depletion 
(\cite{Bacmann02} 2002, 2003). This deuteration of formaldehyde in
pre-stellar cores may occur partially in the CO-depleted gas-phase and partially 
on the surface of dust grains, followed by some inefficient desorption mechanism. 
A further spectacular confirmation of enhanced 
deuteration in CO-depleted gas came from the detection of abundant H$_2$D$^+$, 
likely the most abundant ion, in the prestellar core L1544 (\cite{Caselli03} 2003). 
\cite{Phillips03} (2003) suggested that in CO-depleted gas, even the multiple 
deuterated forms of H$_3^+$ may be abundant and play a role in the molecular 
deuteration enhancement. The suggestion has been fully confirmed by the modelling 
of \cite{Roberts03} (2003), which shows that including HD$_2^+$ and D$_3^+$ in 
the chemical network increases dramatically the molecular deuteration, and allows 
the production of the large atomic D/H ratio predicted by the methanol observations 
(\cite{Parise02} 2002).

In this paper, we report the first detection of triply-deuterated methanol CD$_3$OH 
in space, performed towards the solar-type protostar IRAS~16293. We also present 
a multifrequency observation of $^{13}$CH$_3$OH, used to derive the column density 
of the main isotopomer CH$_3$OH. These observations provide yet another
stringent test to confirm the validity of grain surface models.


\section{Observations and results}\label{sec-obs}

\begin{figure*}[!ht]
\includegraphics[width=16cm,angle=-90]{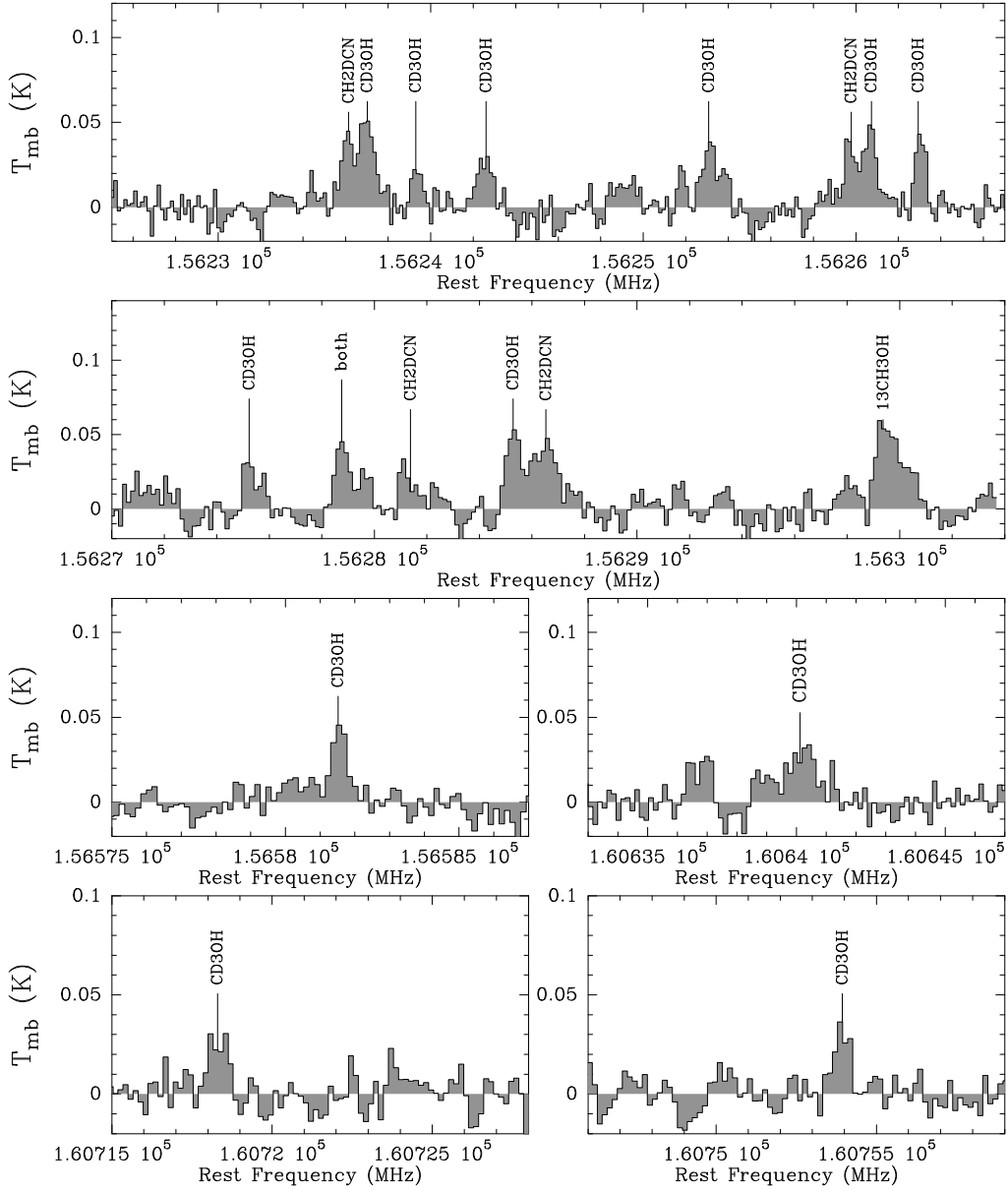}
\caption{CD$_3$OH detected lines. The intensities are reported in main-beam 
brightness temperature. The label ``both'' indicates the blending of one CD$_3$OH 
and one CH$_2$DCN lines. This latter CD$_3$OH line has not been considered in the 
population diagram analysis. }
\label{lines}
\end{figure*}

\begin{table*}[!ht]
\begin{center}
\caption{Main-beam intensities$^1$, peak temperatures$^2$ and widths for the observed 
CD$_3$OH and $^{13}$CH$_3$OH transitions$^3$.  
}
\begin{tabular}{ccccccc}
\hline
\hline
Frequency & Transition & $\mu^2S$ & E$_{up}$ & $\int{{\rm T}_{\rm mb}dv}$ & T$_{\rm mb}$ &  $\Delta$v \\
GHz & & Debye$^2$ & K & K~km~s$^{-1}$ & mK & km~s$^{-1}$\\
\hline
CD$_3$OH & & & & & & \\
\hline
156.237016$^{\star}$ & 4$_1$-3$_1$ E$_2$ & 2.94 & 21.5 & 0.102 $\pm$ 0.023 & 51 & 1.8 $\pm$ 0.4 \\
156.239295 & 4$_2$-3$_2$ A-    & 2.44 & 42.0   & 0.032 $\pm$ 0.012 & 26 & 1.2 $\pm$ 0.5 \\
156.242613 & 4$_0$-3$_0$ A+    & 3.13 & 18.8 & 0.059 $\pm$ 0.018 & 31 & 1.8 $\pm$ 0.5 \\ 
156.253079 & 4$_2$-3$_2$ A+    & 2.44 & 42.0   & 0.056 $\pm$ 0.015 & 38 & 1.4 $\pm$ 0.4 \\  
156.260737$^{\star}$ & 4$_3$-3$_3$ E$_2$ & 1.39 & 55.5 & 0.061 $\pm$ 0.018 & 46 & 1.3 $\pm$ 0.4 \\
156.262936 & 4$_3$-3$_3$ E$_1$ & 1.37 & 46.9 & 0.044 $\pm$ 0.011 & 43 & 1.0 $\pm$ 0.2 \\  
156.275238 & 4$_1$-3$_1$ E$_1$ & 2.93 & 33.1 & 0.034 $\pm$ 0.021 & 26 & 1.2 $\pm$ 1.2 \\  
156.285288$^{\star}$ & 4$_2$-3$_2$ E$_2$ & 2.35 & 36.3 & 0.065 $\pm$ 0.021 & 54 & 1.1 $\pm$ 0.3 \\ 
156.581519 & 8$_1$-7$_0$ E$_2$ & 4.15 & 70.2 & 0.050 $\pm$ 0.010 & 46 & 1.0 $\pm$ 0.2 \\ 
160.640122 & 2$_0$-2$_1$ E$_2$ & 2.38 & 15.95& 0.090 $\pm$ 0.019 & 26 & 3.2 $\pm$ 0.5 \\   
160.718291 & 6$_2$-5$_1$ E$_1$ & 2.22 & 50.2 & 0.039 $\pm$ 0.009 & 29 & 1.3 $\pm$ 0.2 \\  
160.753934 & 1$_0$-1$_1$ E$_2$ & 1.45 & 12.2 & 0.038 $\pm$ 0.009 & 35 & 1.0 $\pm$ 0.2 \\ 
\hline
$^{13}$CH$_3$OH & & & & & & \\
\hline
156.299374 &  5$_{05}$-5$_{-15}$ &   0.697  &  47.1  &   0.19$\pm$0.08 & 65 & 3.0 $\pm$ 0.5 \\
160.507694 &  2$_{12}$-3$_{03}$  &   0.300  &  21.3  &   0.08$\pm$0.02 & 29 & 2.4 $\pm$ 0.6 \\       
330.194042 &  7$_{-17}$-6$_{-16}$ &   5.55   &  69.0  &   0.51$\pm$0.16 & 120 & 3.8 $\pm$ 0.6 \\  
330.252798 &  7$_{07}$-6$_{06}$  &   5.66   &  63.4  &   0.43$\pm$0.15 & 100  & 4.0 $\pm$ 1.2 \\      
330.265233 &  7$_{-61}$-6$_{-60}$&   1.50   & 253.4  &   0.15$\pm$0.10 &  80  & 1.7 $\pm$ 1.0 \\  
330.277270 &  7$_{61}$-6$_{60}$  &   1.50   & 258.1  &   0.12$\pm$0.08 &  60  & 2.0 $\pm$ 0.7 \\
330.277270 &  7$_{62}$-6$_{61}$  &   1.50   & 258.1  &   0.12$\pm$0.08 &  60  & 2.0 $\pm$ 0.7 \\
330.319110 &  7$_{52}$-6$_{51}$  &   2.78   & 202.0  &   0.45$\pm$0.21 &  90  & 4.8 $\pm$ 0.9 \\
330.319110 &  7$_{53}$-6$_{52}$  &   2.78   & 202.0  &   0.45$\pm$0.21 &  90  & 4.8 $\pm$ 0.9 \\  
330.342534 &  7$_{44}$-6$_{43}$  &   3.81   & 144.2  &   0.07$\pm$0.05 &  50  & 1.3 $\pm$ 6.7 \\
330.342534 &  7$_{43}$-6$_{42}$  &   3.81   & 144.2  &   0.07$\pm$0.05 &  50  & 1.3 $\pm$ 6.7 \\  
330.408395 &  7$_{34}$-6$_{33}$  &   4.61   & 111.4  &   0.91$\pm$0.19 & 200  & 4.2 $\pm$ 0.7 \\
330.442421 &  7$_{16}$-6$_{15}$  &   5.69   &  84.49 &   0.42$\pm$0.12 & 140  & 2.8 $\pm$ 0.9 \\
330.535822 &  7$_{25}$-6$_{24}$  &   5.14   &  85.80 &   0.25$\pm$0.11 &  40  & 5.6 $\pm$ 1.2 \\   
330.535890 &  7$_{-26}$-6$_{-25}$&   5.20   &  89.45 &   0.25$\pm$0.11 &  40  & 5.6 $\pm$ 1.2 \\ 
\hline
\end{tabular}
\label{obs}
\end{center}
$^1$The fluxes were derived using Gaussian fits, and the uncertainty given is $\sqrt{\sigma_{stat}^2 + \sigma_{cal}^2}$ where $\sigma_{stat}$ is the statistical error and $\sigma_{cal}$ the calibration uncertainty (15\%).
$^2$The noise rms is 8~mK for the CD$_3$OH data and 37~mK for the $^{13}$CH$_3$OH data.
$^3$A star following the frequency indicates that the line is close to a 
CH$_2$DCN line and was fitted by a two-component Gaussian fit (see text).
\end{table*}

Using the IRAM 30-meter telescope (Pico Veleta, Spain), we 
detected the 12 CD$_3$OH lines reported in Table \ref{obs}.  
The telescope was pointed at the coordinates $\alpha$(2000)~=~16$^h$32$^m$22.6$^s$ 
and $\delta$(2000)~=~-24$^{\circ}$28$'$33.0$''$. The observations were
performed in April 2003. Two receivers were used simultaneously at 2 mm, 
to observe two bands around 156 and 160 GHz, with typical 
system temperatures of about 230 and 250 K respectively. These receivers were connected 
to the VESPA autocorrelator divided in six units. The telescope beam width is 
approximately 15$''$ at 160 GHz. All observations were performed using the wobbler 
switching mode with an OFF position 4$'$ from the source. The pointing accuracy was 
monitored regularly on strong continuum sources, and was found to be better than 3$''$.
All spectra were obtained with an integration time of 750 minutes. The rms noise 
is equal to 8~mK (T$_{mb}$) for a spectral resolution of 0.3~km s$^{-1}$.

Observed spectra are shown in Fig. \ref{lines}. 
The measured intensities, linewidths and main-beam temperatures are reported 
in Table~\ref{obs}. The frequencies of all detected lines have 
previously been measured in the laboratory with an accuracy of 25 kHz 
(\cite{walsh98} 1998), while the transition strengths
and energy levels were estimated from the published spectroscopic constants 
(\cite{Pcross98} 1998) using the methanol program at Ohio State. 

Some of the CD$_3$OH lines (indicated by a star in Table \ref{obs}) are close 
to CH$_2$DCN lines. In that case, the intensity was derived by using a two-component 
Gaussian fit, so the quoted fluxes have a further uncertainty due to the relative 
line contribution. 

Two $^{13}$CH$_3$OH lines at 156~GHz were observed simultaneously to the CD$_3$OH lines. 
In addition, we analysed 330~GHz $^{13}$CH$_3$OH observations obtained using the JCMT 
in January 2000, with an rms noise of 37~mK. The beam size of 
the JCMT is 15$''$ at the considered frequencies, i.e. equivalent to the beam size 
of the 30-meter at 160~GHz. Detailed information concerning the  $^{13}$CH$_3$OH spectra 
is presented in Table \ref{obs}.


\section{Derivation of the column densities}

\begin{figure}
\includegraphics[width=9cm]{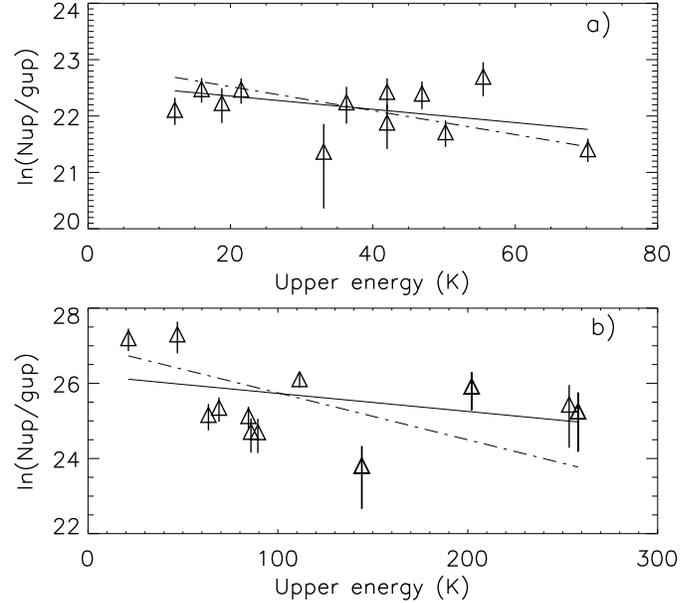}
\caption{Rotational diagram for : {\bf a)} CD$_3$OH (solid line T$_{rot}$\,=\,85 K, dashed line: T$_{rot}$\,=\,47 K, see text), {\bf b)} $^{13}$CH$_3$OH (solid line : T$_{rot}$\,=\,208 K, dashed line : T$_{rot}$\,=\,80 K, see text). 
Column densities are averaged on a 10$''$ beam.}
\label{rotdiag}
\end{figure}

We derived the abundance of CD$_3$OH using the method of rotational diagrams. 
The A and E species are considered to be linked by ion-molecule reactions that transfer 
molecules from one species to the other.
We then computed one single rotational diagram for the two species, presented in 
Fig.~\ref{rotdiag} (a).
We averaged the level column densities on a 10$''$ beam, as in \cite{Parise02} (2002), 
following the suggestion by \cite{vanDishoeck95} (1995) of enhanced methanol emission in the 
central 10$''$ region of IRAS~16293. A more recent study of the spatial distribution
of CH$_3$OH was performed by \cite{Schoier02} (2002), and showed evidence for an abundance
jump of methanol of two orders of magnitude in the inner part of the envelope ($\le$ 150 AU).
However, in the following we consider averaged abundances on a 10$''$ beam for consistency 
with the \cite{Parise02} (2002) study. 

The ground  E state is estimated to lie about 4.6~K above the ground  A state.
An A state has a relative spin-torsional weight of 11 whereas the relative spin-torsional 
weight of an E state is 16.
The partition function was computed from the asymmetric-top approximation :
$$ Z(T)=11\times Z_A(T) + 16\times \exp(-4.6/T) \times Z_E(T) $$
where $Z_A(T)=Z_E(T)=\sqrt{\frac{\pi T^3}{ABC}}$, with A~=~3.3957~K, B~=~0.9529~K and 
C~=~0.9247~K, as determined by \cite{walsh98} (1998).
 
By fitting a straight line to the data in the rotational diagram, we derive 
a rotational temperature of 85\,$\pm$\,28~K, consistent with the rotational 
temperature of CH$_3$OH (\cite{vanDishoeck95} 1995). The CD$_3$OH column density is 
(1.4\,$\pm$\,0.9)$\times$10$^{14}$~cm$^{-2}$. We also derived the CD$_3$OH column density 
by fixing the rotational temperature to the one inferred from the 
CH$_2$DOH and CHD$_2$OH molecules (T$_{rot}$\,=\,47\,$\pm$\,7~K, \cite{Parise02} 2002).
The CD$_3$OH column density is then (7.8\,$\pm$\,2.3)$\times$10$^{13}$~cm$^{-2}$. 
Table~\ref{coldens} lists the column densities for all deuterated methanols observed 
in IRAS~16293. 
 
The column density of $^{13}$CH$_3$OH was derived using the same method, with the molecular 
parameters  taken from the Cologne Database for Molecular Spectroscopy (\cite{Muller01} 2001).
 The rotational diagram is presented in Fig \ref{rotdiag} (b). The inferred rotational 
temperature is 208\,$\pm$\,70~K and the column density averaged over a 10$''$ beam is 
(2.6\,$\pm$\,1.8)$\times$10$^{14}~$cm$^{-2}$. We also computed the column density for fixed
rotational temperatures of 50~K and 80~K. The inferred value is 
(1.4\,$\pm$\,0.6)$\times$10$^{14}~$cm$^{-2}$, 
independent of the temperature in this range. Using the $^{12}$C/$^{13}$C  ratio of 70 
derived by \cite{Boogert00} (2000), we derive a column density 
of (9.8$\,\pm$\,4.2)$\times$10$^{15}$~cm$^{-2}$ for CH$_3$OH. This value is nearly 
3 times higher than the column density used by \cite{Parise02} 
(2002), inferred from CH$_3$OH observations from \cite{vanDishoeck95} (1985). 
The fractionation ratios, relative to this new estimate of the CH$_3$OH column density, 
are reported in Table \ref{coldens} for all deuterated isotopomers of methanol.

\begin{table}[!ht]
\caption{Derived column densities and fractionation ratios relative to CH$_3$OH for deuterated methanols in IRAS~16293}
\begin{tabular}{cccc}
\hline
\hline
Molecule & T$_{rot}$(K) & N (cm$^{-2}$) & fractionation \\
CD$_3$OH & 85 $\pm$ 28 & (1.4 $\pm$ 0.9)$\times$10$^{14}$ & 1.4 $\pm$ 1.4 \% \\
         & 47 $\pm$ 7$^a$  & (7.8 $\pm$ 2.3)$\times$10$^{13}$ & 0.8 $\pm$ 0.6 \%  \\  
CHD$_2$OH$^b$ & 47 $\pm$ 7 & (6.0 $\pm$ 2.2)$\times$10$^{14}$ & 6 $\pm$ 5 \% \\
CH$_2$DOH$^b$ & 48 $\pm$ 3 & (3.0 $\pm$ 0.6)$\times$10$^{15}$ & 30 $\pm$ 20 \% \\
CH$_3$OD$^b$ & 20 $\pm$ 4 & (1.5 $\pm$ 0.7)$\times$10$^{14}$ & 2 $\pm$ 1 \% \\
\hline
\end{tabular}
\label{coldens}
{$^a$fixed temperature, see text. $^b$Observed in Parise et al. 2002}
\end{table}


\section{Discussion and conclusions}

The main result of this Letter is the first detection of triply-deuterated 
methanol in space, with 12 detected transitions.
This discovery follows the detection of doubly-deuterated as well as 
singly-deuterated isotopomers towards the same object (\cite{Parise02} 2002). 
Observations of multiple isotopomers of methanol represent a powerful 
constraining tool for chemical processes that lead to such a high deuteration.

It is interesting to compare these observations to the predictions of the 
simple grain chemistry scheme of \cite{Rodgers02} (2002). If the D atoms are 
randomly distributed in the methanol isotopomers (i.e. this scheme does not 
consider any activation barrier for the reactions but rather assumes that all 
reactions are equiprobable), the fractionation ratios R of each isotopomer 
relative to CH$_3$OH should scale as follows: 
R(CH$_3$OD)\,=\,$\alpha$, R(CH$_2$DOH)\,=\,3$\alpha$,
R(CHD$_2$OH)\,=\,3$\alpha^2$ and 
R(CD$_3$OH)\,=\,$\alpha^3$, 
where $\alpha$ is the accreting atomic D over H ratio. The three independent 
observations of CH$_2$DOH, CHD$_2$OH and CD$_3$OH 
are consistent within the error bars with a value of 0.1$-$0.2 for the D over H 
accretion rate. Accounting for the different mass of the atoms,
this ratio corresponds to an abundance ratio in the gas-phase of 
D/H\,=\,$\sqrt{2}\times$(0.1$-$0.2)\,=\,0.15$-$0.3. However, this simple scheme fails to 
explain the observed low abundance of CH$_3$OD.

\begin{figure}
\includegraphics[width=9cm]{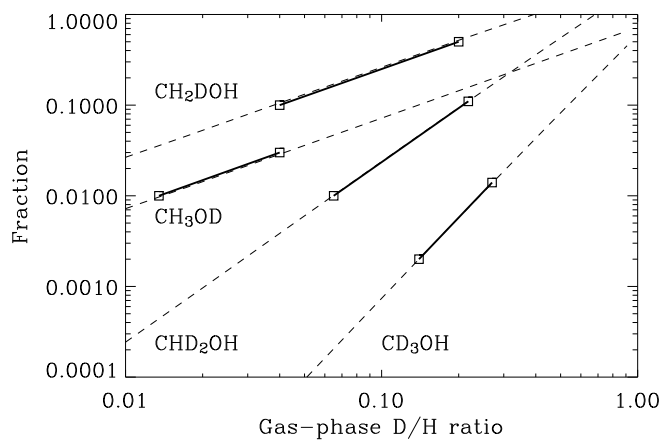}
\caption{The deuterium fractionation ratio for the various deuterated isotopomers 
of methanol is plotted against the abundance ratio of deuterium to hydrogen atoms in the 
gas phase. Dashed lines: model results of  Stantcheva et al. 2003. Thick lines: 
observations of IRAS~16293. }
\label{model}
\end{figure}

More accurate grain chemical models accounting for different activation barriers for the 
reactions have been developed in the last few years. We compare in the following our
observations with the model developed by \cite{Stantcheva03} (2003). This model is based 
on the direct solution of the master equation and therefore gives essentially 
the same predictions
as the Monte Carlo models described by \cite{Caselli02} (2002) or \cite{Charnley97} (1997). 
 Fig.~\ref{model} shows predictions for fractionation ratios of 
deuterated isotopomers of methanol relative to CH$_3$OH versus the atomic 
D/H ratio in the accreting gas (\cite{Stantcheva03} 2003) when
the various isotopomers are formed by active grain chemistry. In the limit of low 
temperature (10~K), this model essentially gives the ratios corresponding to a random 
distribution of deuterium atoms. 
Observed fractionation ratios with their error bars have been overlaid on each curve, 
allowing the derivation of the required atomic D/H ratio in the gas-phase. 
The CD$_3$OH, CHD$_2$OH and CH$_2$DOH observations are consistent with a  formation on 
grain surfaces with an atomic D/H abundance ratio of 0.1$-$0.2. Such a high atomic 
fractionation ratio in the gas phase is predicted by the recent gas-phase model of 
\cite{Roberts03} (2003), which involves not only H$_2$D$^+$ but also D$_2$H$^+$ 
and D$_{3}^{+}$ as precursors for deuterium fractionation, when 
the density of gas is very high and heavy species such as CO are strongly depleted.
 
As can be seen in Fig.~\ref{model}, CH$_3$OD appears to be under-deuterated when 
compared with the grain chemical predictions. It is possible that the CH$_3$OD fractionation
may be affected in the warm gas; e.g., this isotopomer  
may be preferentially converted into CH$_3$OH when released in the gas-phase 
by protonation reactions 
followed by dissociative recombination with an electron (\cite{Charnley97} 1997, 
\cite{Parise02} 2002) :
\vspace{-0.2cm}
\begin{center}
$ {\rm CH_3OD + H^+_3 \rightarrow  CH_3ODH^+ + H_2}$

${\rm  CH_3OHD^+ + e^- \rightarrow  CH_3OH + D .} $
\end{center}
\vspace{-0.2cm}
The corresponding reactions with H$_2$D$^+$, HD$_2^+$ and D$_3^+$ come of little 
importance in view of their low abundance in the warm gas of the hot core.
This hypothesis, which assumes that protonation 
reactions attack the oxygen end of the methanol only (Osamura, Roberts \& Herbst, in prep), 
could be tested by observing the CH$_2$DOD isotopomer. This observation may be 
difficult due to the expected low intensity of the lines.

\begin{acknowledgements}

E. Herbst acknowledges the support of the National Science
Foundation (US) for his research program in astrochemistry.

\end{acknowledgements}

{}


\end{document}